\documentclass[superscriptaddress,reprint,pra]{revtex4-1}
\usepackage{graphics}
\usepackage{bm}
\usepackage{bbm}
\usepackage{amsmath}
\usepackage{amssymb}
\usepackage{graphicx}
\usepackage{amsthm}
\begin{document}
\title{Fermionic entanglement in the Lipkin model}
\author{M.\ Di Tullio}
\affiliation{IFLP/CONICET and Departamento  de F\'{\i}sica,
    Universidad Nacional de La Plata, C.C. 67, La Plata (1900), Argentina}
\author{R.\ Rossignoli}
\affiliation{IFLP/CONICET and Departamento  de F\'{\i}sica,
    Universidad Nacional de La Plata, C.C. 67, La Plata (1900), Argentina}
    \affiliation{Comisi\'on de Investigaciones Cient\'{\i}ficas (CIC), La Plata (1900), Argentina}
\author{M.\ Cerezo}
\affiliation{Theoretical Division, Los Alamos National Laboratory, Los Alamos, NM 87545, USA}
\author{N.\ Gigena}
\affiliation{IFLP/CONICET and Departamento  de F\'{\i}sica,
    Universidad Nacional de La Plata, C.C. 67, La Plata (1900), Argentina}

\begin{abstract}
We examine the fermionic  entanglement in the ground state of the fermionic Lipkin model and its relation with bipartite entanglement. It is first shown that 
the one-body entanglement entropy, which quantifies the minimum distance to a fermionic Gaussian state, 
behaves similarly to the mean-field order parameter and is essentially proportional to the total bipartite entanglement between the upper and lower modes, a quantity meaningful only in the fermionic realization of the model. We also analyze the entanglement of the reduced state of four single-particle modes (two up-down pairs), showing that its fermionic concurrence is strongly peaked at the phase transition and behaves 
differently from the corresponding up-down entanglement. 
We finally show that the first measures and the up-down reduced entanglement can be correctly described through a basic mean-field approach supplemented with symmetry restoration, whereas the concurrence requires at least the inclusion of RPA-type correlations for a proper prediction. 
Fermionic separability is also discussed.
 \end{abstract}
\maketitle
\section{Introduction}

Quantum correlations have attracted much attention ever since the theory of quantum mechanics was introduced \cite{EPR.1935,Schr.35}. In particular, quantum entanglement has become a central concept in present-day physics, due to its role as a resource in quantum information science \cite{Be.93,Sch1,NC.00,JL.03,VG.03} and the deep insight it provides in many-body physics  \cite{ON.02,LK.03,AF.08,VCM.08,ECP.10,DC.18} and fundamental problems \cite{PSW.06,VV.13,QT.15}.  Nonetheless, the definition of quantum entanglement relies heavily on the tensor product structure of the Hilbert space and the subsequent notion of locality \cite{VR.05}.  
When considering systems of indistinguishable particles, the standard definition cannot be directly applied, thus preventing a  straightforward extension of well-known measures.

In the last years, there has been a great interest in  generalizing the notion of entanglement to fermionic systems \cite{SC.01,SL.01,ES.02,WV.03,GM.04,KW.09,ZP.02,S.03,FL.13,BF.14,PV.14,IV.13,OK.13,SG.14,GR.15,GR.17,MB.16,DGR.18,DD.16,AG.17,ID.17,SH.19}. Two different approaches have been taken: \textit{mode entanglement} \cite{ZP.02,S.03,FL.13,BF.14,PV.14}, where the subsystems are described as a collection of single-particle (sp) modes (with the fermion number of each subsystem not necessarily fixed) and entanglement depends on the chosen basis for the whole sp space;   
and \textit{particle entanglement} \cite{SC.01,SL.01,ES.02,WV.03,GM.04,KW.09,IV.13,OK.13,SG.14,GR.15,GR.17,MB.16,DGR.18}, where entanglement is basis independent and defined beyond  antisymmetrization, i.e., as a vanishing quantity in the case of a Slater Determinant (SD).  In the case of two fermions, a Schmidt-like decomposition of a general pure state in terms of SD's can be performed \cite{SC.01},  and this entanglement  can be  associated with that of two distinguishable fermions, in the sense of occupying orthogonal subspaces \cite{SC.01,SL.01,ES.02,WV.03}. 

In \cite{GR.15} an entropic measure of mode entanglement was introduced, which, after minimization over all  sp bases,  becomes  a measure of fermionic entanglement, in the sense of vanishing iff the state is a SD. The ensuing quantity is based upon the one-body density matrix and  represents essentially a minimum distance to a fermionic Gaussian state \cite{DGR.18}. 
It can be extended to states with no fixed fermion number (but fixed number parity) through the quasiparticle density matrix, vanishing iff the state is a quasiparticle vacuum \cite{GR.15}. These measures can also be applied to mixed states through convex roof extension. In a sp space of dimension 4 (first nontrivial case) they  can be evaluated by means of a fermionic concurrence \cite{SC.01,GR.15},  analogous to the standard concurrence \cite{Wo.97}, vanishing iff the state is a convex combination of fermionic Gaussian states. 

The aim of this work is to examine these measures in the well-known  Lipkin model, considering the original fermionic version  \cite{LM.651,LM.652,LM.653}. This model can be exactly solved  through its mapping onto a spin system and has been extensively used as a benchmark for testing many-body approximations and studying quantum phase transitions and dynamics \cite{LM.651,LM.652,LM.653,RS.80,GF.78,RC.97,VPA.04,HC.06,CF.08,DC.15}. It has also been employed for analyzing entanglement in its spin realization \cite{VPM.04,LO.05,DV.05,BDV.06,MRC0.08,CU.08,MRC.10,DV.11,DV.12}. Its simulation with circuit QED was considered in  \cite{La.10}. 

We first analyze the one-body entanglement entropy in the  exact ground state (GS) for a finite size. We show that this quantity exhibits a close correlation  with the mean-field (MF) order parameter, becoming significant  in the symmetry breaking MF phase and  saturating for strong couplings. We also show that 
it is approximately proportional to the bipartite mode  entanglement entropy between all up and down fermionic modes, a quantity physically accessible only in the fermionic realization of the model. We then analyze the   entanglement of the reduced state of four fermionic modes (two up-down pairs).  Its fermionic concurrence,   which measures the deviation of the reduced state  from a mixture of SD's, is  explicitly evaluated. It is shown to  correspond to the spin-pair concurrence,  
exhibiting a peak at the MF GS transition and being strongly affected by the coupling anisotropy. In contrast,  the up-down mode entanglement of the reduced state, which  can be measured through the pertinent negativity,   behaves similarly to previous global measures. The  analytic description of these  quantities  through different approximations, including standard and symmetry-restored  MF approaches as well as RPA (random-phase approximation)-based schemes are also provided.  The isotropic limit and the fermionic GS separability are as well discussed.

The formalism is briefly reviewed in section \ref{I}, while the model and its fermionic entanglement are analyzed in \ref{II}. Conclusions are finally provided in \ref{III}. 

 \section{Formalism}
 \label{I}
We consider an $n$-dimensional single-particle (sp) Hilbert space $\mathcal{H}$ spanned by the fermionic operators $c^\dag_i$, $i=1,\dots,n$, satisfying the anticommutation relations $\{c_i,c_j^\dag\}=\delta_{ij}$, $\{c_i,c_j\}=\{c_i^\dag,c_j^\dag\}=0$. 
Given a pure fermionic state $|\Psi\rangle$ (assumed to have a definite number  of fermions: 
$\sum_i c^\dag_i c_i|\Psi\rangle=N|\Psi\rangle$),  the elements of the  one-body density matrix are   $\rho_{ij}^{\rm sp}=\langle c_j^\dag c_i\rangle\equiv\langle\Psi| c_j^\dag c_i|\Psi\rangle$. 
In \cite{GR.15} we defined the one-body entanglement entropy as 
\begin{eqnarray}
E(|\Psi\rangle)&=&{\rm Tr}\, h(\rho^{\rm sp})\nonumber\\
&=&-\sum_k\, f_k\log_2 f_k+(1-f_k)\log_2(1-f_k) \,,\label{1}
\end{eqnarray}
where $h(f)=-f\log_2 f-(1-f)\log_2(1-f)$ and $f_k=\langle a^\dagger_k a_k\rangle$ denote the eigenvalues of $\rho^{\rm sp}$ ($a_k=\sum_i U_{ki}c_i$, with $U$ unitary satisfying $\langle a^\dag_l a_k\rangle=(U\rho U^\dag)_{kl}=f_k\delta_{kl}$). Eq.\ (\ref{1}) is  proportional to the minimum, over all sp bases of $\mathcal{H}$, of the average entanglement entropy between a single fermionic mode and its orthogonal complement \cite{GR.15}: 
\begin{equation}
E(|\Psi\rangle)=\mathop{\rm Min}_{\{c_i\}}\sum_i h(\langle c^\dagger_i c_i\rangle)\,.\label{min}
\end{equation}
This quantity is a measure of fermionic entanglement, in the sense that it vanishes iff $|\Psi\rangle$ is a Slater Determinant (SD), i.e.\ iff  $(\rho^{\rm sp})^2=\rho^{\rm sp}$ ($f_k=0$ or $1$ $\forall$ $k$), and remains invariant under one-body unitary transformations $|\Psi\rangle \rightarrow \exp[-ic^\dag O c]|\Psi\rangle$ (with $c^\dag Oc=\sum_{i,j}O_{ij}c^\dag_i c_j$ an hermitian one-body operator), which just lead to a unitary transformation of $\rho^{\rm sp}$ 
($\rho^{\rm sp}\rightarrow e^{-iO}\rho^{\rm sp}e^{iO}$).

As shown in \cite{DGR.18}, Eq.\ (\ref{1}) can also be interpreted as the {\it minimum  relative entropy} (in the grand  canonical ensemble) between $\rho=|\Psi\rangle\langle\Psi|$ and {\it any} fermionic Gaussian state $\rho'$: 
\begin{equation}
E(|\Psi\rangle)=\mathop{\rm Min}_{\rho'}S(\rho||\rho')\,.\label{min2}
\end{equation}
Here $S(\rho||\rho')={\rm Tr}\,\rho(\log_2\rho-\log_2\rho')$ is the quantum relative entropy \cite{Wh.78,VRMP.02} satisfying $S(\rho||\rho')\geq 0$, with $S(\rho||\rho')=0$ iff  $\rho'=\rho$, and $\rho'$ 
 is the exponent of a one-body operator: 
\begin{equation}
\rho'=\exp[-\lambda_0-c^\dag\Lambda c]\,,
\end{equation}
with $\lambda_0=\ln\,{\rm Tr}\,\exp[-c^\dag\Lambda c]=\sum_k \ln (1+e^{-\lambda_k})$ and $\lambda_k$ the eigenvalues of the matrix $\Lambda$. Therefore, Eq.\ (\ref{1})  represents a measure of the distance between $\rho$ and its closest fermionic Gaussian state, thus vanishing whenever $|\Psi\rangle$ is a SD. The formalism can be extended to pure states $|\Psi\rangle$ with no fixed fermion number but fixed number parity $P_N=\exp[-i\pi \sum_i c^\dagger_i c_i]$ \cite{GR.15,DGR.18}. 

In the case of a two-fermion state, which can be always written by means of the fermionic Schmidt decomposition \cite{SC.01,SL.01,ES.02}  as $|\Psi\rangle=\sum_{k=1}^{n_s} \sqrt{\lambda_k}c^\dag_k c^\dag_{\overline{k}}|0\rangle$, with $\lambda_k>0$, $\sum_k \lambda_k=1$ and $n_s$ the Slater rank 
($k,\overline{k}$ denote orthogonal sp states), it is easily seen that 
$f_k=f_{\overline{k}}=\lambda_k$ and  $E(|\Psi\rangle)=2\sum_k h(\lambda_k)>0$ iff  $n_s>1$, 
i.e., iff $|\Psi\rangle$ is not a SD. The state can then be viewed as an entangled state of two  fermions occupying orthogonal subspaces, spanned by the sp states $k$ and $\bar{k}$ respectively, with $\frac{1}{2}E(|\Psi\rangle)$ a measure of the associated bipartite  entanglement. 

Let us consider now a subset of sp states spanning  a subspace ${\cal H}_A$ of  the sp space ${\cal H}$, with ${\cal H}_B$ its orthogonal complement (${\cal H}_A\oplus {\cal H}_B={\cal H}$). A general state can be written as  $|\Psi\rangle=\sum_{\mu,\nu}C_{\mu\nu}|\mu\nu\rangle$, where   $|\mu\nu\rangle=[\prod_{i\in A}(c_i^\dag)^{n_i^{(\nu)}}][\prod_{j\in B}(c_j^\dag)^{n_j^{(\mu)}}]|0\rangle$ are SDs and  $n_i^{(\nu)}=0,1$  the occupation number of sp state $i$ in configuration $\nu$. The reduced state of ${\cal H}_A$ can be obtained by taking the partial trace, $\rho_A=\sum_{\mu,\mu'}(CC^\dag)_{\mu\mu'}|\mu\rangle\langle \mu'|$, such that $\langle \Psi|O_A|\Psi\rangle={\rm Tr}_A\,\rho_A O_A$ for any observable $O_A$ built with  fermion operators acting just on ${\cal H}_A$.  Its entropy $S(\rho_A)=-{\rm Tr}\rho_A\log_2\rho_A$ measures the bipartite mode entanglement between this set and its orthogonal complement. For instance, if $\mathcal{H}_A$ contains  just one sp level $i$, 
\begin{equation}
    \rho_A=\begin{pmatrix}
    \langle c_i^\dag c_i\rangle & 0\\
    0 & \langle c_i c_i^\dag\rangle
    \end{pmatrix}\,,
\end{equation}
in the basis $\{c^\dag_i|0\rangle,|0\rangle\}$, with $\langle c_ic^\dag_i\rangle=1-\langle c^\dag_i c_i\rangle$, and $S(\rho_A)=h(\langle c_i^\dag c_i\rangle)$ is the mode entanglement entropy between $i$ and all remaining sp states, used in (\ref{min}). 

The  convex-roof extension of the one-body entropy (\ref{1}) for a mixed state $\rho_A$ (with $[\rho_A,\sum_{i\in A} c^\dag_i c_i]=0$) is   
\begin{equation} E(\rho_A)=\mathop{\rm Min}_{\sum_{\alpha}q_\alpha |\psi_\alpha^A\rangle\langle\psi_\alpha^A|=\rho_A}
E(|\psi^A_\alpha\rangle)\label{Eform}\,,\end{equation}
where $q_\alpha>0$, $\sum_\alpha q_\alpha=1$ and minimization is over all decompositions of $\rho_A$ as convex combinations of pure states $|\psi_\alpha^A\rangle$ (with definite particle number) 
\cite{ES.02,GR.15}. This quantity,  analogous to  the bipartite entanglement of formation \cite{BDSW.96}, is a measure of fermionic entanglement for mixed states in the sense of  vanishing iff $\rho_A$ is a convex mixture of SD's,   remaining  invariant under one-body transformations $\rho_A\rightarrow e^{-ic^\dag Oc}\rho_A e^{ic^\dag O c}$. Moreover, if $|\Psi\rangle$ is a SD, $E(\rho_A)=0$ for any subspace ${\cal H}_A$  since, due to the validity of Wick's theorem,  $\rho_A$ will be a Gaussian state \cite{DGR.18}.  

In order to have $E(\rho_A)>0$,  a  subspace ${\cal H}_A$ of dimension $\geq 4$ is required, since any pure state with fixed particle number in a subspace of lower dimension is always a SD \cite{SC.01,GR.15}. And for ${\cal H}_A$ of dimension 4, $E(\rho_A)>0$ ensures finite biparitte entanglement of $\rho_A$ for  {\it any} bipartition of ${\cal H}_A$ into subspaces of finite dimension (${\cal H}_A={\cal H}_{A_1}\oplus{\cal H}_{A_2}$) \cite{GR.17,DGR.18}. Moreover, in this case, an analytic evaluation of (\ref{Eform}) for {\it any} state $\rho_A$ having fixed number parity $P_N$ becomes feasible through the fermionic concurrence $C(\rho_A)$  \cite{SC.01,ES.02,GR.15}, like in a two-qubit system \cite{Wo.97}.  The result is 
$E(\rho_A)=2[h(f_+)+h(f_-)]=4h(f_+)$ \cite{GR.15}, with $f_{\pm}=(1\pm\sqrt{1-C^2(\rho_{A})})/2$ and 
\begin{equation}
C(\rho_{A})={\rm Max}[2\lambda_{\rm max}-{\rm Tr}\,R(\rho_{A}),0]\,,\label{Cf}
\end{equation}
where $\lambda_{\rm max}$ is the largest eigenvalue of 
$R(\rho_A)=\sqrt{\rho_A^{1/2}\tilde{\rho}_A\rho_A^{1/2}}$  and $\tilde{\rho}_A=T\rho_A^*T$. We will consider in what follows states $\rho_A$ with even number parity $P_N=+1$. In this case a four dimensional ${\cal H}_A$ leads to an $8$-dimensional Fock space with $P_N=+1$ and the operator $T$ is represented, in the basis $\{|0\rangle,c_1^\dag c_2^\dag|0\rangle,c_1^\dag c_3^\dag|0\rangle,c_1^\dag c_4^\dag|0\rangle,-c^\dag_1 c^\dag_2 c^\dag_3 c^\dag_4|0\rangle, 
  c^\dag_3 c^\dag_4|0\rangle,-c^\dag_2c^\dag_4|0\rangle,\\c^\dag_2c^\dag_3|0\rangle\}$, by the matrix 
\begin{equation}
    T=\begin{pmatrix}
    0 &\mathbbm{1}_4\\
    \mathbbm{1}_4 & 0
    \end{pmatrix}\,,
\end{equation}
with $\mathbbm{1}_4$ the $4\times 4$ identity. For instance, any pure two-fermion state in such ${\cal H}_A$ can be writtten as  $|\psi_A\rangle=(\alpha a^\dag_1 a^\dag_2+\beta a^\dag_3a^\dag_4)|0\rangle$ \cite{Sch1,GR.15}, with    $|\alpha|^2+|\beta|^2=1$, in which case, for $\rho_A=|\psi_A\rangle\langle\psi_A|$, 
Eq.\ (\ref{Cf}) leads to $C(\rho_A)=2|\alpha\beta|$ and hence to  $f_{+}=|\alpha|^2$, $f_-=|\beta|^2$, which are the (two-fold degenerate) eigenvalues of $\rho^{\rm sp}$. 

\section{Lipkin Model}
\label{II}
The original fermionic Lipkin model \cite{LM.651,LM.652,LM.653} describes a system of fermions in  $2\Omega$ sp states $|p\pm\rangle$,  $p=1,\ldots,\Omega$, with energies $\varepsilon_{p\pm}=\pm \varepsilon/2$, fully connected through  uniform two-body couplings. The Hamiltonian is  
\begin{eqnarray}
\nonumber H&=&\tfrac{1}{2}\varepsilon\sum_{p}
(c_{p+}^{\dagger} c_{p+}-c_{p-}^{\dagger} c_{p-})-W\sum_{p,q} c_{p+}^{\dagger}c_{q-}^{\dagger}c_{q+} c_{p_-}\nonumber\\&& -\tfrac{1}{2}V\sum_{p,q}(c_{p+}^{\dagger}c_{q+}^{\dagger}c_{q-} c_{p_-}+c^\dagger_{p-}c^\dagger_{q-}c_{q+}c_{p+})\,,
\label{hamlip}
\end{eqnarray}
where the $V$-coupling creates (and destroys) particle-hole pairs over  the unperturbed  GS 
\begin{equation}|\Psi_0\rangle=(\prod_p c^\dagger_{p-})|0\rangle\label{psi0}\,,\end{equation}
while the $W$-coupling is a hopping term for these pairs. 
 
 As is well known \cite{LM.651,LM.652,LM.653,RS.80}, this Hamiltonian can be mapped into a spin system  through the operators 
 \begin{eqnarray}
S_{\pm}&=&S_x{\pm}i S_y= \sum_p  c_{p{\pm}}^{\dagger }c_{p{\mp}}\,,\nonumber\\
S_z&=&\frac{1}{2} \sum_p\left( c_{p+}^{\dagger }c_{p+}-c_{p-}^{\dagger }c_{p-} \right)\,,\label{spinop1}\end{eqnarray}
which  satisfy exact $SU(2)$ commutation relationships: $[S_+,S_-]=2S_z$,  $[S_z,S_{\pm}]=\pm S_{\pm}$. The Hamiltonian (\ref{hamlip}) can  then be rewritten as 
\begin{eqnarray}
\hspace{-1cm}H&=&\varepsilon S_z -\tfrac{W}{2}(S_{+}S_-+S_{-}S_+-N)- \tfrac{V}{2}(S_+^2+S_-^2)\label{H1}\\
&=& \varepsilon  S_z-V_x\left(S_x^2 + \chi S_y^2\right)+V_x\tfrac{1+\chi}{4}N\,,\label{H2}
\end{eqnarray}
where $V_x=W+V$, $\chi=\frac{W-V}{W+V}$ and $N=\sum_{p,\mu=\pm}c^\dagger_{p\mu}c_{p\mu}$ is the  fermion number operator.  In what follows we will consider the half-filled case of $N=\Omega$ fermions, where  the total spin $S$ can reach  the maximum value $\Omega/2$. We will set, without loss of generality, $|\chi|\leq 1$ (i.e.\ $V_y\leq V_x$)  and  focus on the attractive case $V_x\geq 0$.  These two conditions imply $W\geq 0$, $V\geq 0$.  It also convenient to scale the coupling strength as 
\begin{equation}V_x=v_x/(\Omega-1)\,,\label{vx}\end{equation}
such that $\langle H\rangle/\Omega$ stays finite and $\Omega$-independent for large $\Omega$ at fixed $v_x,\chi$. And at the mean-field level  (see sec.\ \ref{IIB}) the GS transition to the symmetry-breaking phase will take place  exactly  at $v_x=\varepsilon$ $\forall$ $\Omega$. 

 The Hamiltonian $H$ obviously commutes with the total spin $S^2=S_x^2+S_y^2+S_z^2$. In the  half-filled attractive case the minimum energy is reached for maximum spin $S=\Omega/2$, i.e., for the completely symmetric multiplet. The exact GS $|\Psi\rangle$ can then be obtained by diagonalizing the equivalent spin-hamiltonian (\ref{H2}) within the $S=\Omega/2$ subspace, of dimension $\Omega+1$, which contains the unperturbed GS (\ref{psi0}) and where there is always just one fermion per site $p$ (note that $[H,n_p]=0$, 
with $n_p=\sum_{\mu=\pm}c^\dagger_{p\mu}c_{p\mu}$the number of fermions at site $p$). It will then have the form 
 \begin{eqnarray}
    |\Psi\rangle &=& \sum_{K=0}^{\Omega} C_K |K\rangle\,,\label{GSX}
\end{eqnarray}
where $K$ denotes the number of fermions in the upper level and  $|K\rangle\equiv|S=\Omega/2,S_z=K-\Omega/2\rangle\propto S_{+}^{K}|\Psi_0\rangle$ 
are the eigenstates of $S_z$ for maximum total spin $S=\Omega/2$. All  coefficients $C_K$ will be real and of the same sign if $V>0$.   

Moreover, $H$ also commutes with the {\it $S_z$-parity $P_z$}:   
\begin{equation}[H,P_z]=0\,,\;\;P_z=
\exp[i\pi (S_z+\Omega/2)]\,,\end{equation}
which for $N=\Omega$  is equivalent to the parity 
of the number of fermions in the upper level: $P_z=\exp[i\pi \sum_p c^\dagger_{p+}c_{p+}]$. This 
implies that in any non-degenerate GS the sum in (\ref{GSX}) will run  over either even or odd values of $K$, according to the GS parity $P_z=\pm 1$. This symmetry will have a deep influence in fermionic entanglement measures. 

In the isotropic case  $\chi=1$ ($V=0$), the Hamiltonian  (\ref{H2}) also commutes with $S_z$, implying that the sum in (\ref{GSX}) reduces to a single term.  The exact eigenenergies for $N=\Omega$  and $\chi=1$ become, using Eq.\ (\ref{H2}),  
\begin{equation}
    E_{SK}={\textstyle\varepsilon(K-\frac{\Omega}{2})-V_x[S(S+1)-(K-\frac{\Omega}{2})^2-\frac{\Omega}{2}]}
    \,,\label{ESM}
\end{equation}
with $S\leq \Omega/2$  and $\frac{\Omega}{2}-S\leq K
\leq \frac{\Omega}{2}+S$. It is then verified that for $V_x>0$, the GS   is  obtained for  maximum spin $S$,  with $K$ starting at $0$ for $V_x=0$ and  ending up at $[\Omega/2]$ for large $V_x$. 
Thus, the $\chi=1$ GS undergoes $[\Omega/2]$ transitions $K\rightarrow K+1$ as $V_x$ increases from $0$, at the  couplings \begin{equation}V_x^{K}=\frac{\varepsilon}
{\Omega-1-2K}\,,\;\;K=0,\ldots,[{\textstyle\frac{\Omega}{2}}]-1\,,
\label{VxK}\end{equation} 
with $v_x^0=V_x^0/(\Omega-1)=\varepsilon$ coinciding with the MF critical value. 
For fixed $\chi\in(0,1)$ ($0<V<W$) and finite $\Omega$, these transitions will evolve into  $[\Omega/2]$ $S_z$-parity transitions.  On the other hand, for $\chi\in[-1,0]$ ($V\geq W$) the exact GS  will have even parity $P_z=+1$  $\forall$ $V_x>0$, owing  to the dominant role of the $V$-coupling. 

\subsection{Exact GS entanglement}
\subsubsection{One body entropy}
 We start with the evaluation of the one-body entanglement entropy (\ref{1}). Due to  conservation of the single site fermion number $n_p$, the elements of the one-body density matrix $\rho^{\rm sp}$ in a state of the form (\ref{GSX}) satisfy $\langle c^\dagger_{p\mu}c_{q\nu}\rangle=\delta_{pq}\langle c^\dag_{p\mu} c_{p\nu}\rangle$. And 
 due to translational invariance over the states $p$, they form $\Omega$ identical blocks 
\begin{equation}
    \rho^{\rm sp}_p =\begin{pmatrix}\langle c^\dagger_{p+}c_{p+}\rangle&\langle c^\dag_{p-}c_{p+}\rangle\\\langle c^\dag_{p+}c_{p-}\rangle&\langle c^\dag_{p-}c_{p-}\rangle\end{pmatrix}=
    \begin{pmatrix} f_+ &0\\0 &f_- \end{pmatrix}\,, \label{rhospx}
\end{equation}
 where   
$\langle c^\dag_{p\pm}c_{p\mp}\rangle=\langle S_{\pm}\rangle/\Omega=0$ owing to $P_z$  conservation and $f_{\pm}=\left(\tfrac{1}{2}\pm\langle S_z\rangle/\Omega\right)$, i.e., $f_+=\langle K\rangle/\Omega=1-f_-$  
with $\langle K\rangle=\sum_K C_K^2 K$. 
The one-body  entropy $E\equiv  E(|\Psi\rangle)$  then becomes  
\begin{equation}
    E = \Omega[h(f_+)+h(f_-)]=-2\Omega (f_+\log_2 f_++f_-\log_2 f_-)\,.\label{Ssp}
\end{equation}
It is determined just by $\langle S_z\rangle$,  vanishing only if $|\langle S_z\rangle|=\Omega/2$. This  shows that  {\it any} state (\ref{GSX}) with {\it definite parity $P_z$} is not a SD unless 
 $C_{K}=\delta_{K0}$ or $\delta_{K\Omega}$. This includes in particular the states $|K\rangle$ (GS's for $\chi=1$),  where  $f_+=K/\Omega\equiv k$ and 
\begin{equation}
    E_K = -2\Omega [k\log_2 k+(1-k)\log_2(1-k)]\,,\label{Sspk}
\end{equation}
is positive for $1\leq K\leq \Omega-1$, reflecting the fact that they are collective excitations  generated by $S_+$ and are hence not SD's. Eq.\ (\ref{Ssp}) is $2\Omega$ times the single spin entanglement entropy in the spin picture (i.e.\ the entanglement of spin $i$ with the rest of the system),  where the block (\ref{rhospx}) represents the reduced single spin density matrix. 
\begin{figure}
{\includegraphics[scale=0.8]{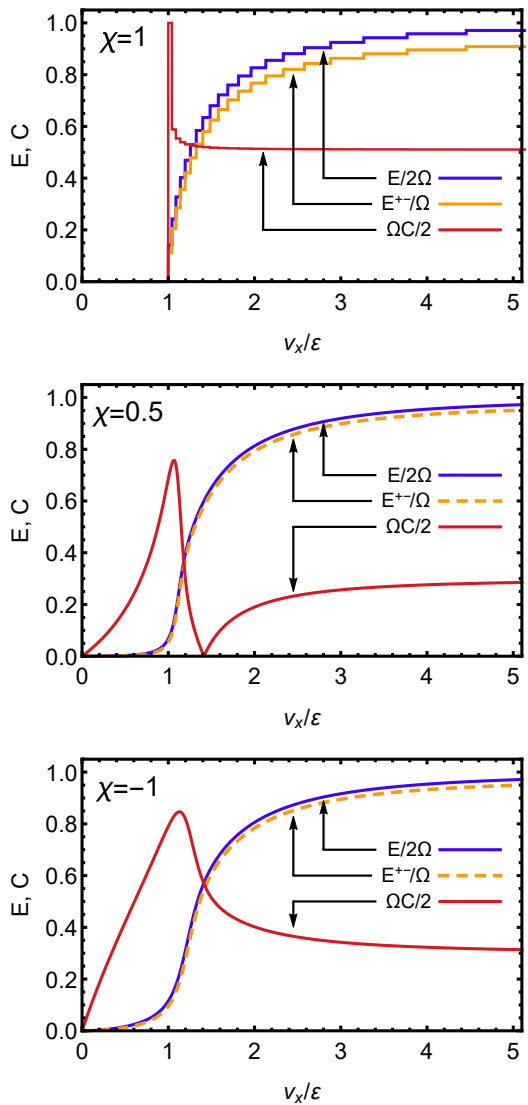}}
    \caption{The intensive one-body entropy $E/(2\Omega)$, Eq.\ (\ref{Ssp}), together with the intensive up-down entanglement entropy $E^{+-}/\Omega$  and the scaled fermionic concurrence $\Omega C/2$, in the GS of Hamiltonian (\ref{hamlip})--(\ref{H2})  as a function of the relative coupling strength $v_x/\varepsilon$, for three different anisotropies $\chi$ and $\Omega=50$ fermions. 
        All labels are dimensionless.} 
    \label{fig1}
\end{figure}

The behavior of the one-body entropy (\ref{Ssp})  as a function of $v_x/\varepsilon$ is shown in  Fig.\ \ref{fig1} for $\Omega=50$ 
and different anisotropies $\chi$, together with other entanglement measures discussed below. It becomes significant  for $v_x>\varepsilon$, i.e.\ in the strongly coupled regime where the MF GS exhibits $S_z$-parity breaking (see sec.\ \ref{IIB}), increasing monotonously with increasing strength $v_x/\varepsilon$ and saturating for $v_x/\varepsilon\rightarrow\infty$. Moreover, it is almost independent of $\chi$,  except for the  steps visible in the $\chi=1$ case which reflect the $S_z$ transitions $K\rightarrow K+1$ (the steps  due to the $P_z$ transitions at $\chi=0.5$ are already very small at the size shown and are not appreciable). An analytic description of $E$ for large $\Omega$ is provided in \ref{IIB}.

\subsubsection{Up-down entanglement entropy} 
The fermionic version of the model enables to  consider a bipartition where Alice has access to the $\Omega$ lower single fermion levels and Bob to the $\Omega$ upper levels, which  has no physical counterpart in a pure spin realization. It is only in the fermionic case where this partition becomes  meaningful and the ensuing up-down mode entanglement  can be considered as a physical resource. 
Let us also remark that  due to the definite  $S_z$-parity of the GS, 
each side will contain fermionic states with definite number parity ($P_{N_\pm}=P_z$ ($\pm P_z$) for $\Omega$ even (odd)), entailing that  
arbitrary linear combinations of the involved local states will not break the number parity superselection rule \cite{FL.13,FL.15}. 

In order to compute this entanglement  exactly, we first note that each state $|K\rangle$ can be explicitly written as an equally weighted sum of $\binom{\Omega}{K}$  orthogonal  SD's with $K$ fermions up  and $\Omega-K$ fermions down at different sites, 
\begin{equation}
    |K\rangle=\frac{1}{\sqrt{\binom{\Omega}{K}}}
    \sum_\alpha \left(\prod_p (c^\dagger_{p+}c_{p-})^{n_{p\alpha}}\right)|\Psi_0\rangle\,,
    \label{obe}
\end{equation}
where $n_{p\alpha}=0,1$, $\sum_p n_{p\alpha}=K$ and $\alpha=1,\ldots,\binom{\Omega}{K}$.  Since all  terms in the previous sum involve different orthogonal states at each side, Eq.\ (\ref{obe})  constitutes the Schmidt decomposition \cite{NC.00} for the up-down partition. Then, the up-down entanglement entropy of the states  $|K\rangle$ is 
\begin{equation}E^{+-}_K=\log_2 \binom{\Omega}{K}\,.\label{Sudk}\end{equation} 
Using Eq.\ (\ref{obe}), the ensuing up-down entanglement entropy determined by the exact GS  (\ref{GSX})  is  
\begin{eqnarray}
    E^{+-}&=&
    \sum_K C_K^2[\log_2{\textstyle\binom{\Omega}{K}}-\log_2 C_K^2]
    \label{Sud2}\,,
\end{eqnarray}
where the first sum in (\ref{Sud2}) represents the average entanglement stemming from the states $|K\rangle$  while the second is that emerging from the distribution over these states.

While $E^{+-}$ does not require in principle a correlated GS 
 (it can be non-zero in a SD, see sec.\ \ref{IIB}), it is here essentially  {\it half} the one-body entropy (\ref{Ssp})  for sufficiently large $\Omega$:  In  the states  $|K\rangle$, an expansion of (\ref{Sudk}) for large $\Omega$ and $K$, with $k=K/\Omega$ finite,  leads to 
\begin{eqnarray}E^{+-}_K&\approx &-\Omega[k\log_2 k+(1-k)\log_2 (1-k)]\nonumber\\&&-\log_2\sqrt{2\pi\Omega k(1-k)}\nonumber\\&=&E_K/2+O(\log_2\Omega)\,.\label{Sudka}\end{eqnarray}
A similar relation $E^{+-}\approx E/2$ holds  for large $\Omega$ in a typical definite parity GS (\ref{GSX})  (see sec.\ \ref{IIB}).  This result indicates a close correlation between the fermionic entanglement measured by $E(|\Psi\rangle)$ and the bipartite up-down mode entanglement in the exact definite parity GS. This enables an approximate estimation of the latter, which can be  considered as a quantum resource, through an easily accessible one-body average ($\langle S_z\rangle$). 
The behavior of $E^{+-}$ with $v_x/\varepsilon$ is also depicted in Fig.\ \ref{fig1}, where the approximate proportionality is verified.

\subsubsection{Fermionic concurrence and reduced up-down entanglement\label{concur}}
We now examine the fermionic and up-down entanglement  in the reduced state $\rho_{pq}$ of four single fermion states $p_{\pm}$, $q_{\pm}$, $p\neq q$, which is  a mixed state for $\Omega>2$ and is here the first non-trivial case for both measures.  We first note that  since $n_p=1$, the reduced state $\rho_p$ of a single pair of modes $p_\pm$ (represented in principle by a $4\times 4$ matrix in the basis $\{|0\rangle,c^\dag_{p+}|0\rangle,c^\dag_{p-}|0\rangle,c^\dag_{p+}c^\dag_{p-}|0\rangle$) contains here just  a diagonal $2\times 2$ non-zero block $\rho_p$ identical with Eq.\ (\ref{rhospx}) in the restricted basis  $\{c^\dag_{p+}|0\rangle,c^\dag_{p-}|0\rangle\}$: 
\begin{equation} \rho_p=\begin{pmatrix}
\langle c^\dag_{p+}c_{p+}c_{p-}c^\dag_{p-}\rangle&\langle c^\dag_{p-}c_{p+}\rangle\\
\langle c^\dag_{p+}c_{p-}\rangle&\langle c_{p+}c^\dagger_{p+}
c^\dag_{p-}c_{p-}\rangle 
\end{pmatrix}=\begin{pmatrix}f_+&0\\0&f_-\end{pmatrix}\,.\label{rps}\end{equation}
This state has obviously no fermionic entanglement, in the sense that it is just a mixture of elementary single fermion states   ($\rho_p=\sum_{\nu=\pm} f_\nu c^\dag_{p\nu} |0\rangle\langle 0|c_{p\nu}$). Besides, it has no up-down mode entanglement either, since $P_z$-conservation implies $\langle c^\dagger_{p\pm}c_{p\mp}\rangle=0$  and  prevents coherence between both single fermion states (i.e., terms $\propto c^\dag_{p+}|0\rangle\langle 0|c_{p-}$). Thus, if Alice and Bob have access just to a single $p$ state 
($p-$ for Alice, $p+$ for Bob) the joint state $\rho_p$ contains just classical correlations (it is equivalent to a two-qubit state $f_-|01\rangle\langle 01|+f_+|10\rangle\langle 10|$, with $|01\rangle=c^\dag_{p-}|0\rangle$, $|10\rangle=c^\dag_{p+}|0\rangle$).  

\begin{figure}
\includegraphics[scale=1.]{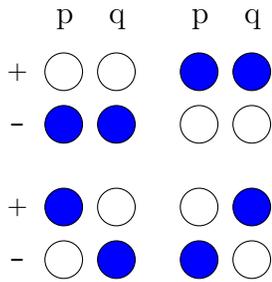}
\caption{The possible occupancies of four sp states $p_{\pm}$, $q_{\pm}$ in the GS of the Hamiltonian (\ref{hamlip}). The reduced state (\ref{rpq}) contains coherences between the two upper states and, separately, between the two lower states, which lead to a nonzero fermionic concurrence and up-down entanglement.} 
\label{fig2}
\end{figure}

The situation changes when they have access to  two different states $p\neq q$ ($p-,q-$ for Alice, $p+,q+$ for Bob). Since the number $n_p$ of fermions per site $p$ is $1$, the support of $\rho_{pq}$ (in principle a $16\times 16$ matrix)  will just involve the four two-fermion states 
$\{c_{p+}^\dagger c_{q+}^\dagger|0\rangle,c_{p+}^\dagger c_{q-}^\dagger|0\rangle,c_{p-}^\dagger c_{q+}^\dagger|0\rangle,c_{p-}^\dagger c_{q-}^\dagger|0\rangle\}$ (see Fig.\ \ref{fig2}). In this restricted basis it will be given by 
\begin{equation}
   \rho_{pq} = \begin{pmatrix}
    \langle n_{p+}n_{q+}\rangle& 0 & 0 & \langle s_{p-}s_{q-}\rangle \\ 
    0 & \langle n_{p+}n_{q-}\rangle &\langle s_{p-}s_{q+}\rangle & 0\\\\
    0 & \langle s_{p+}s_{q-}\rangle& \langle n_{p-}n_{q+}\rangle&0\\
     \langle s_{p+} s_{q+}\rangle & 0 & 0 &  \langle n_{p-}n_{q-}\rangle 
    \end{pmatrix}\label{rpq}
\end{equation}
where $n_{p\pm}=c^\dagger_{p\pm} c_{p\pm}$ and $s_{p\pm}=c^\dagger_{p\pm}c_{p\mp}$. 
In the GS (\ref{GSX}),  its elements are independent of $p,q$ and can be exactly evaluated in terms of global averages:   
\begin{eqnarray}
\langle s_{p\pm}s_{q\pm}\rangle&=&\frac{\langle S_\pm^2\rangle}{\Omega(\Omega-1)}\,,\label{rr14}\\
\langle s_{p\pm}s_{q\mp}\rangle&=&
\frac{\Omega^2/4-\langle S_z^2\rangle}{\Omega(\Omega-1)} =\langle n_{p\pm}n_{q\mp}\rangle\,,\label{rr22}\\
\langle n_{p\pm}n_{q\pm}\rangle&=&\left(\frac{1}{4}\pm\frac{\langle S_z\rangle}{\Omega}+\frac{\langle S_z^2\rangle-\Omega/4}{\Omega(\Omega-1)}\right)\label{rr11}\,.
\end{eqnarray}

The fermionic entanglement of this mixed state, which indicates its deviation from a statistical mixture of SD's, can be measured through the fermionic concurrence (\ref{Cf}), which becomes  
\begin{eqnarray}
C&=&2{\rm Max}[|\langle s_{p+}s_{q+}\rangle|-\langle n_{p+}n_{q-}\rangle,\nonumber\\&&|\langle s_{p+}s_{q-}\rangle|-\sqrt{\langle n_{p+}n_{q+}\rangle\langle n_{p-}n_{q-}\rangle},0]\,.\label{CC}
\end{eqnarray}
This concurrence is  equivalent to that of a two-qubit system, coinciding with that of a spin pair in the spin realization \cite{VPM.04,DV.05,MRC0.08} (i.e., with that measuring bipartite mode entanglement between $p\pm$ and $q\pm$ states). It can be regarded as  ``parallel'' (``antiparallel'')  when  the first (second) term in (\ref{CC}) is positive, and scales  as $\Omega^{-1}$ 
\cite{VPM.04,MRC0.08}.   

\begin{figure}
    \centering
    \includegraphics[scale=0.9]{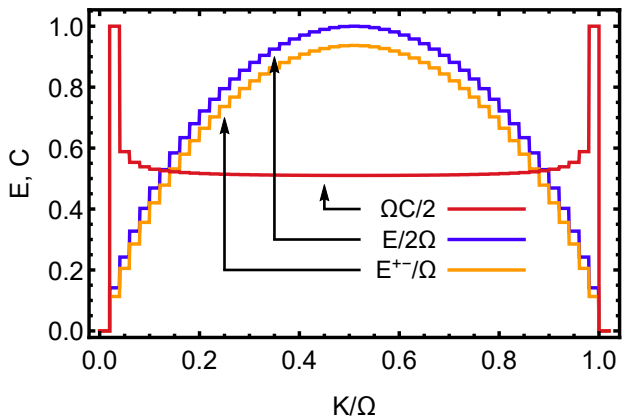}
    \caption{Fermionic entanglement measures in the states $|K\rangle=|S=\frac{\Omega}{2},M=K-\frac{\Omega}{2}\rangle$ for $\Omega=50$ fermions. Here $E$ is the one-body  entanglement entropy (\ref{Sspk}),  $E^{+-}$ the up-down entanglement entropy (\ref{Sudk}) and $C$ the  concurrence (\ref{CK}).} 
    \label{fig3}
\end{figure}

As seen in Fig.\ \ref{fig1}, the behavior of $C$  differs from that of previous entanglement measures. 
It is  peaked at $v_x\approx \varepsilon$, i.e.\ at the onset of the symmetry-breaking MF phase,  and it is strongly affected by the anisotropy. For $\chi=1$ the GS's are the states $|K\rangle$ and the concurrence is then only of antiparallel type,  as $\langle S_\pm^2\rangle=0$, and  given by 
\begin{equation}
    C_K=\frac{2}{\Omega}\frac{1}{1+\sqrt{1-\frac{\Omega-1}{K(\Omega-K)}}}\,,\;\;1\leq K\leq \Omega-1\,.\label{CK}
\end{equation}
vanishing for $K=0,\Omega$ and satisfying $C_K=C_{\Omega-K}$. It is sharply peaked  at  $K=1$ (and $K=\Omega-1$),  with $C_1=2/\Omega$ (the maximum attainable value in this model \cite{MRC0.08}), dropping  significantly already for $K=2$ ($C_2/C_1\approx 2-\sqrt{2}\approx 0.586$ for large $\Omega$)  and reaching its  minimum at $K=\Omega/2$ ($C_{\Omega/2}=\frac{1}{\Omega-1}$). 
An explicit comparison between the concurrence $C_K$ and the entanglement entropies $E_K$ and $E^{+-}_K$ is shown in Fig.\ \ref{fig3}. 

The concurrence remains peaked at  $v_x\approx \varepsilon$ for other anisotropies, but the  sharpness of the peak decreases as $\chi$ decreases, as seen in Fig.\ \ref{fig1}. In addition, for  $\chi\in(0,1)$ (central panel) 
it vanishes at the {\it separability  point} 
\begin{equation}v_x=\varepsilon/\sqrt{\chi}\label{fac}\,,\end{equation} 
(see Appendix) where it changes from antiparallel to parallel \cite{RVF.04,RCM.08}. This point corresponds to the first GS $S_z$-parity transition, where it becomes two-fold degenerate and the GS subspace is spanned by two non-orthogonal SDs (see sec.\ \ref{IIB}). For $\chi<0$ GS separability  no longer occurs and $C$ is positive and parallel $\forall$ $v_x>0$. 
 
\begin{figure}
    \centering
    \includegraphics[scale=0.9]{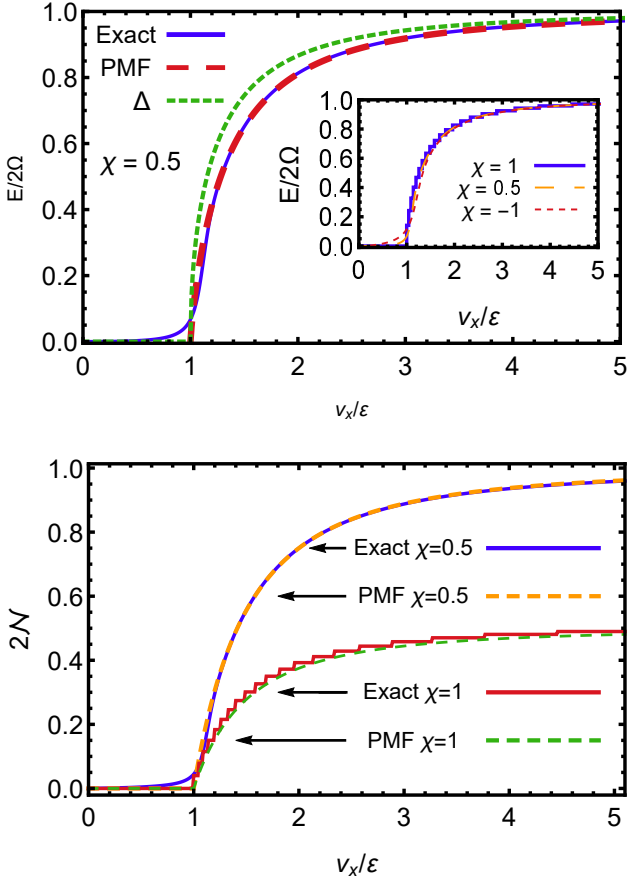}
    \caption{Top panel: Comparison  between the exact one-body entanglement entropy and the value obtained from the mean-field approximation through  basic $S_z$-parity projection (PMF), Eq.\ (\ref{Ssptheta}), for $\chi=0.5$ and  $\Omega=50$ fermions. The MF order parameter $\Delta=2|\langle s_x\rangle|_{\rm mf}=\sin\theta$ is also depicted. The inset depicts the exact result for different values of $\chi$, showing the weak dependence on the anisotropy.  Bottom: The exact up-down negativity in the reduced state $\rho_{pq}$, Eq.\ (\ref{neg}), for $\chi=0.5$ and $\chi=1$ (Eq.\ \ref{negk}), together with the mean-field results
    obtained through basic $S_z$-parity ($\chi=0.5$) and $S_z$ ($\chi=1$) projections, Eqs.\ (\ref{neg2})-- (\ref{neg3}), which are almost coincident with the exact ones.} 
    \label{fig4}
\end{figure}

The  finite  fermionic concurrence of the reduced state $\rho_{pq}$ warrants non-zero bipartite entanglement for {\it any} bipartition of the four dimensional sp space  \cite{DGR.18}. In particular, $\rho_{pq}$ will lead to a finite up-down mode entanglement, 
which can be quantified through the pertinent negativity \cite{VW.02,ZHSL.99,SH.19}. This partition involves four distinct states at each side: $|0\rangle$, $c^\dagger_{p-}c^\dagger_{q-}|0\rangle$, $c^\dagger_{p-}|0\rangle$ and $c^\dagger_{q-}|0\rangle$ for Alice and similar states at the upper level for Bob (Fig.\ \ref{fig2}), leading to a two-qudit system with $d=4$. $P_z$ symmetry implies that just states with the same local fermion number parity are connected in $\rho_{pq}$, entailing that partial trasposition will not mix local states with different number parity and standard formulas can be applied \cite{SH.19}.  The ensuing negativity is just minus the sum of the two negative eigenvalues of the partial trasposed matrix:
\begin{equation}
    \mathcal{N}^{+-}=|\langle s_{p+}s_{q+}\rangle| + |\langle s_{p+}s_{q-}\rangle|\,.\label{neg}
\end{equation}
As seen in Fig.\ \ref{fig4} (bottom panel), the behavior of this quantity resembles that of the global entropies  $E$ and  $E^{+-}$, increasing monotonously for increasing $v_x/\varepsilon$.  While still weakly dependent on $\chi$ 
for $\chi<1$, it is reduced by half for $\chi\rightarrow 1$ : 
In the states $|K\rangle$ the first term  in (\ref{neg}) vanishes and Eq.\ (\ref{rr22}) leads to ($k=K/\Omega$)  \begin{equation}
    \mathcal{N}^{+-}_K=\frac{\Omega}{\Omega-1}k(1-k)\,,
    \label{negk}
\end{equation}
 approaching $\frac{1}{4}\frac{\Omega}{\Omega-1}$ for $v_x/\varepsilon\rightarrow \infty$. However, for $\chi<1$ both terms in (\ref{neg}) contribute and the ensuing negativity as a function of $v_x/\varepsilon$ becomes essentially twice the value for $\chi=1$, as verified in Fig.\ \ref{fig4}  (see section \ref{IIB}).  
 
  Previous considerations hold of course for $\Omega>2$. We remark that in the trivial $\Omega=2$ case,  $\rho_{pq}$ becomes pure ($\rho_{pq}=|\Psi\rangle\langle\Psi|$) and {\it all} previous quantities become {\it equivalent}:  $C=2{\cal N}^{+-}=2\sqrt{f_+(1-f_+)}$, and $E=4E^{+-}=4h(f_+)$, 
  with $f_+=|\beta|^2$ for a $P_z=+1$ GS  
  \[ |\Psi_+\rangle=\alpha |K=0\rangle+\beta|K=2\rangle=(\alpha c^\dagger_{p-}c^\dagger_{q-}+\beta c^\dagger_{p+}c^\dagger_{q+})|0\rangle\]
  ($|\alpha|^2+|\beta|^2=1$), and $f_+=1/2$ for a $P_z=-1$ GS 
  \[|\Psi_-\rangle=|K=1\rangle=\frac{1}{\sqrt{2}}(c^\dagger_{p+}c^\dagger_{q-}+c^\dagger_{p-}c^\dagger_{q+})|0\rangle\] 
where all previous quantities are maximum.

\subsection{Approximate description\label{IIB}}
We now discuss the evaluation of previous  measures through approximate methods, with the aim of identifying their main sources and  obtaining an analytic description, exact in the  thermodynamic limit  $\Omega\rightarrow\infty$. 

\subsubsection{Mean-field approach\label{MFA}}

 We start with the basic mean-field (MF) approach, where the GS is approximated by a SD, here of the form  
 \begin{equation}|\Psi_{\rm mf}\rangle=e^{-i\theta\bm{n}\cdot\bm{S}}|\Psi_0\rangle=\prod_{p}{c'}^\dagger_{p-}|0\rangle\,,\label{mfs}\end{equation}
where $\bm{S}=(S_x,S_y,S_z)$,  ${c'}^\dagger_{p\pm}=e^{-i\theta\bm{n}\cdot\bm{S}}c^\dagger_{p\pm}e^{i\theta\bm{n}\cdot\bm{S}}$ are rotated fermion operators and $\bm{n}$ a unit vector. Both $\theta$ and $\bm{n}$ are to be obtained from the minimization of the mean energy, which by means of Wick's theorem becomes 
\begin{equation}\langle \Psi_{\rm mf}|H|\Psi_{\rm mf}\rangle=\Omega[\varepsilon\langle s_z\rangle-
v_x(\langle s_x\rangle^2+\chi\langle s_y\rangle^2)]\,,\label{emf}\end{equation} 
where $s_\mu=S_\mu/\Omega$. For $|\chi|<1$, the minimum  will be always obtained for 
$\langle \bm{s}\rangle$ lying in the $x,z$ plane ($\bm{n}.\bm{S}=S_y$), which leads to $\langle \bm{s}\rangle=-\frac{1}{2}(\sin\theta,0,\cos\theta)$ and 
\begin{equation}{c'}_{p\pm}^\dag=
 \cos\tfrac{\theta}{2}\,c_{p\pm}^\dag \pm \sin \tfrac{\theta}{2}\,c_{p\mp}^\dag\,.\label{cp}\end{equation}
Minimization of (\ref{emf}) with respect to $\theta$ yields 
\begin{equation}\cos\theta=\left\{\begin{array}{ll}1\,,&v_x\leq\varepsilon\\\varepsilon/v_x\,,&v_x>\varepsilon\end{array}\right.\,,\label{ct}\end{equation}
i.e., to a normal phase with $\theta=0$ for  $v_x\leq\varepsilon$ and to an {\it $S_z$-parity breaking phase} with 
$|\theta|\in(0,\pi/2)$ (and hence $\Delta\equiv 2|\langle s_x\rangle|=\sin\theta>0$) for $v_x>\varepsilon$. In the latter the sign of $\theta$ remains arbitrary, leading to a two-fold degeneracy of the MF GS (for $\chi=1$, this degeneracy becomes continuous, as the orientation of $\langle\bm{s}\rangle$ in the $xy$ plane becomes  arbitrary). Note that $\theta$ is independent of $\chi$ (for $|\chi|\leq 1$). 

Any SD implies a zero one-body entropy and fermionic concurrence. In particular, the one-body density matrix $\rho^{\rm sp}_{\rm mf}(\theta)$ determined by (\ref{mfs}) consists of $\Omega$ identical blocks 
\begin{equation}
    \rho_{p}^{\rm sp}(\theta) =\begin{pmatrix}\langle c^\dagger_{p+}c_{p+}\rangle&\langle c^\dag_{p-}c_{p+}\rangle\\\langle c^\dag_{p+}c_{p-}\rangle&\langle c^\dag_{p-}c_{-}\rangle\end{pmatrix}=
    \frac{1}{2}\begin{pmatrix} 1-\cos\theta & \sin \theta\\ \sin\theta & 1+\cos\theta
    \end{pmatrix}\,, \label{mft}
\end{equation}
whose eigenvalues are obviously $1$ and $0$. 
Nonetheless,  it is possible to extract an effective MF  one-body entanglement entropy  by considering just the diagonal terms in (\ref{mft}), since the exact  off-diagonal terms vanish due to the $S_z$-parity symmetry. This can be justified through a basic $S_z$-parity restoration (see next subsection), which leads to the  one-body density $(\rho_{\rm mf}^{\rm sp}(\theta)+\rho^{\rm sp}_{\rm mf}(-\theta))/2$, composed of diagonal blocks 
\begin{equation}\frac{1}{2}(\rho^{\rm sp}_p(\theta)+\rho^{\rm sp}(-\theta))=
\frac{1}{2}\begin{pmatrix} 1-\cos\theta & 0\\ 0 & 1+\cos\theta
    \end{pmatrix}\,.\label{rhoth}\end{equation}
The ensuing one-body entanglement entropy  becomes    
\begin{equation} E_{\rm mf}=-2\Omega\sum_{\nu=\pm}{\frac{1+\nu\cos\theta}{2}\log_2\frac{1+\nu\cos\theta}{2}}\label{Ssptheta}\,, \end{equation}
with $\cos\theta$ given by (\ref{ct}), and is  positive for $\theta\in(0,\pi)$ ($v_x>\varepsilon$), saturating for $v_x/\varepsilon\rightarrow\infty$ ($E\rightarrow 2\Omega$). 

As seen in the top panel of Fig.\ \ref{fig4}, Eqs.\ \eqref{ct}--\eqref{Ssptheta} provide an excellent estimation of the exact one-body entropy  $E$ $\forall$  $v_x>\varepsilon$ if $\Omega$ is not too small,  becoming exact 
for $\Omega\rightarrow\infty$. As previously mentioned and as verified in the inset,  the exact $E$ exhibits only a very weak dependence on the anisotropy $\chi$.  And for $\chi\rightarrow 1$ 
and large $\Omega$, the exact Eq.\ (\ref{VxK}) implies $\langle s_z\rangle=K/\Omega-\frac{1}{2}=-\frac{1}{2}\varepsilon/v_x$ plus terms $O(\Omega^{-1})$,  which is  the MF result $\langle s_z\rangle=-\frac{1}{2}\cos\theta$, with Eq.\ (\ref{Sspk}) becoming identical to \ (\ref{Ssptheta})   ($k=\frac{1}{2}(1-\varepsilon/v_x)$).  Thus, in this model the MF approach is able to provide, through basic $S_z$-parity projection (extraction of diagonal terms) the exact result of the intensive one-body entropy $E/(2\Omega)$  for large $\Omega$. 

Besides, the up-down entanglement entropy directly determined by the MF state (\ref{mfs}) is non-zero: This  state  can be written in the form (\ref{GSX}) with coefficients \begin{equation}C_K(\theta)=\sqrt{\binom{\Omega}{K}}\cos^{\Omega-K}\tfrac{\theta}{2}\,\sin^K\tfrac{\theta}{2}\,,
\label{CKt}\end{equation} 
for $\bm{n}\cdot\bm{S}=S_y$.  Eq.\ (\ref{Sud2}) then leads to 
\begin{equation}
    E^{+-}_{\rm mf}=
    -\Omega\sum_{\nu=\pm}{\frac{1+\nu\cos\theta}{2}\log_2\frac{1+\nu\cos\theta}{2}}=
    \frac{1}{2} E_{\rm mf}\,, \label{mfss}
\end{equation}
which is in agreement with the exact trend  for large $\Omega$  and provides a good approximation in this limit.
The result  (\ref{mfss}) is also apparent as  the up or down occupations at each site $p$ in the MF state (\ref{mfs})  are independent from those at other sites. Hence, the MF result  (\ref{mfss}) is that of $\Omega$ independent  two-qubit systems in a  pure entangled state ${c'}^\dagger_{p-}|0\rangle=
\cos\frac{\theta}{2}|01\rangle+\sin\frac{\theta}{2}|10\rangle$, where $|01\rangle=c^\dagger_{p-}|0\rangle$, $|10\rangle=c^\dagger_{p+}|0\rangle$, which is the single $p$ reduced state at the MF level.  

Notice, however, that the up-down entanglement of a single $p$ subsystem is a MF effect arising from $S_z$-parity breaking. As we have seen in \eqref{rps}, the single $p$ reduced state derived from the exact GS  is mixed and just classically correlated, instead of pure and entangled. Thus, in contrast with MF,  in the exact GS $E^{+-}$ emerges from two-body correlations and accordingly, is not strictly extensive (see also Eq.\ (\ref{pmfsup}) in next section). 

On the other hand, the fermionic four-state  concurrence (\ref{CC}) lies strictly beyond the basic MF approach. Without $S_z$-parity projection, the state (\ref{mfs}) obviously leads to a pure  reduced state $\rho_{pq}(\theta)=|\psi_{pq}\rangle\langle\psi_{pq}|$ with $|\psi_{pq}\rangle={c'}^\dag_{p-}{c'}^\dag_{q-}|0\rangle$ a SD, implying $C=0$. But even  after a basic parity projection, which leads to 
\begin{eqnarray}  \rho_{pq} &=& \tfrac{1}{2}(\rho_{pq}(\theta)+\rho_{pq}(-\theta))\nonumber\\
 &=&\frac{1}{4}\begin{pmatrix}
    (1-\cos\theta)^2& 0 & 0 &  
    \sin^2\theta \\ 
    0 & \sin^2\theta &\sin^2\theta & 0\\\\
    0 & \sin^2\theta&\sin^2\theta&0\\
     \sin^2\theta& 0 & 0 &  (1+\cos\theta)^2
    \end{pmatrix}\label{rpq2}\,,
\end{eqnarray}
we still obtain  $C=0$ $\forall$ $\theta$ (according to Eq.\ (\ref{CC})), 
since this state is a convex combination of Gaussian states.  

Eq.\ (\ref{rpq2}) does lead, however, to a correct estimation of the up-down negativity (\ref{neg}) for $\chi<1$,
\begin{equation}
    {\cal N}^{+-}_{\rm mf}=\frac{1}{2}\sin^2\theta\label{neg2}\;\;\;\;(\chi<1)\,,\end{equation}
which provides an excellent description  for large   $\Omega$, as seen in the bottom panel of Fig.\ \ref{fig4}. Notice  that (\ref{neg2}) is smaller than the result derived directly from the unprojected MF reduced state $\rho_{pq}(\theta)$ (i.e., from the pure state $|\psi_{pq}\rangle$), which leads to ${\cal N}^{+-}_{\rm mf}(\theta)=|\sin\theta|(1+\frac{1}{2}|\sin\theta|)$, since the latter includes terms 
connecting states with different $P_z$ which are removed by the basic projection. 

Eq.\ (\ref{neg2})  fails only in the isotropic limit $\chi\rightarrow 1$, where an $S_z$ projection would be required. The main effect of such projection is just to cancel the term $\langle s_{p+}s_{q+}\rangle=\frac{1}{4}\sin^2\theta$ in (\ref{rpq2}), leaving $\langle s_{p+}s_{q-}\rangle$ essentially unaltered. This leads to half the value (\ref{neg2}): 
\begin{equation}{\cal N}_{\rm mf}^{+-}=\frac{1}{4}\sin^2\theta\;\;\;\;(\chi=1)\,,\label{neg3}\end{equation}  which is in excellent agreement with the exact result, as also seen in Fig.\ \ref{fig4}. In fact, for large $\Omega$ Eq.\  (\ref{VxK}) implies  $k=\frac{1}{2}(1-v_x/\varepsilon)$ and the exact result (\ref{negk}) becomes identical with Eq.\ (\ref{neg3}). 

We finally remark that for $\chi\in(0,1)$, both  MF states $|\Psi_{\rm mf}(\pm\theta)\rangle$ become {\it exact} GS's for {\it any} finite $\Omega$ at the separability  point (\ref{fac}) (see Appendix).  However, the exact limit at this point of the one-body entropy is still given by Eq.\ (\ref{Ssptheta}) for large $\Omega$, and is hence {\it non-zero}, since the exact GS side-limits have definite $P_z$ (see next section).  
 
\subsubsection{$S_z$-Parity projected mean-field\label{pmf}}
A more rigorous extraction of fermionic entanglement measures 
at the MF level can be achieved by considering the exact $S_z$-parity restored states
\begin{equation}
|\Psi^\pm\rangle=\frac{|\Psi_{\rm mf}(\theta)\rangle
\pm|\Psi_{\rm mf}(-\theta)\rangle}{\sqrt{2(1\pm\cos^\Omega\theta)}}
\label{factorpar0}\,,
\end{equation}
where $|\Psi_{\rm mf}(\theta)\rangle$ denotes the state (\ref{mfs}) (for $\bm{n}\cdot\bm{S}=S_y$) and $|\Psi_{\rm mf}(-\theta)\rangle=P_z|\Psi_{\rm mf}(\theta)\rangle$. If the overlap $\langle\Psi_{\rm mf}(-\theta)|\Psi_{\rm mf}(\theta)\rangle=\cos^\Omega\theta$ and other terms of similar order are neglected (they are very small for not too small $\Omega$ and $\theta$), the one-body density $\rho^{\rm sp}$ and the reduced state $\rho_{pq}$ derived from the states (\ref{factorpar0}) are  precisely given by Eqs.\ (\ref{rhoth}) and (\ref{rpq2}). 

Nonetheless, 
it is also possible to derive the exact projected expressions. The contractions derived from the states (\ref{factorpar0}) are  $\langle c^\dagger_{p\mu}c_{q\nu}\rangle_{\pm}=\delta_{pq}\delta_{\mu\nu}f^\pm_\mu$ for $\mu,\nu=\pm$,  where $f^\pm_\nu$ are the projected average occupations  \begin{equation}f^{\pm}_{\nu}=
\frac{1-\nu\cos\theta}{2}\left(\frac{1\mp\nu  \cos^{\Omega-1}\theta}{1\pm\cos^\Omega\theta}\right)\,,
\end{equation}
satisfying  $f^\pm_++f^\pm_-=1$. Hence, these states lead to a diagonal one-body density matrix with $\Omega$ identical blocks 
\begin{equation}
\rho_p^{\pm}=\begin{pmatrix}f_{+}^\pm&0\\0&f^{\pm}_-\end{pmatrix}\,.\label{rhopmf}\end{equation}
Since the difference between $f_\nu^\pm$ and  the unprojected average occupations $\frac{1-\nu \cos\theta}{2}$ of Eq.\ (\ref{rhoth}) is $O(\cos^{\Omega-1}\theta)$, the exact one-body  entropy determined by the states (\ref{factorpar0}), \begin{equation}E_{\pm}=-2\Omega(f_+^{\pm}\log_2f_+^{\pm}+
f_-^{\pm}\log_2f_-^{\pm})\,,\label{SspP}\end{equation}
is very close to the diagonal MF result  (\ref{Ssptheta}) if $\Omega$ and $\theta$ are not too small. 

The up-down entanglement determined by the states (\ref{factorpar0}) can also be exactly evaluated: The ensuing normalized coefficients in the expansion (\ref{GSX}) are just 
\begin{equation}C_K^{\pm}(\theta)=\sqrt{\tfrac{1\pm(-1)^K}{1\pm\cos^\Omega\theta}}\,C_{K}(\theta)\,,\end{equation}
with $C_K(\theta)$ the MF coefficients  (\ref{CKt}), and  lead to 
\begin{equation}E^{+-}_{\pm}=-\Omega \left(\sum_{\nu=\pm}f_\nu^{\pm}\log_2\tfrac{1+\nu\cos\theta}{2}\right) -\log_2 \tfrac{2}{1\pm\cos^\Omega\theta}\,.\label{pmfsup}\end{equation}
Even if terms $O(\cos^\Omega\theta)$  are neglected,  a reduction $\approx-\log_2 2=-1$ is now obtained in comparison with (\ref{mfss}) due to $S_z$-parity projection, which implies a spread over just half the total number of states. This lowering explains the small difference with half the one-body entropy observed in the exact results of Fig.\ \ref{fig1} for  $\chi<1$.  In particular, the up-down entanglement is maximum for $\theta=\pi/2$ ($v_x/\varepsilon\rightarrow \infty)$, in which case (\ref{pmfsup}) yields  $E^{+-}_{\pm}=\Omega-1$, while (\ref{mfss}) leads to $E^{+-}=\Omega$.  An alternative interpretation of this reduction is  that while in the  nonprojected MF, Alice (lower  levels) and Bob (upper levels) share  $\Omega$ independent maximally entangled qubit pairs, in the  projected state the last pair is not independent as its state becomes determined by the parity constraint, thus lowering $E^{+-}$  in one unit. On the other hand, full $S_z$ projection is required for obtaining an accurate evaluation at $\chi=1$, which would lead to the exact result (\ref{Sudk}).

The exact reduced four-mode state $\rho_{pq}$ derived from the states (\ref{factorpar0}) has the form (\ref{rpq}) with  elements ($\nu=\pm$)
\begin{eqnarray}
     \langle s_{p\nu}s_{q\nu}\rangle_{\pm}&=&
      \frac{\sin^2\theta}{4}\left(\frac{1\pm   \cos^{\Omega-2}\theta}{1\pm\cos^\Omega\theta}\right)\nonumber\\
      \langle s_{p\nu}s_{q-\nu}\rangle_{\pm}&=&
      \frac{\sin^2\theta}{4}\left(\frac{1\mp   \cos^{\Omega-2}\theta}{1\pm\cos^\Omega\theta}\right)=\langle n_{p\nu}n_{q-\nu}\rangle_{\pm}\label{elc}\\
          \langle n_{p\nu}n_{q\nu}\rangle_{\pm}&=&
     \frac{1}{2}-\langle s_{p\nu}s_{q-\nu}\rangle_{\pm}\mp\frac{\cos\theta}{2}
          \left(\frac{1\pm   \cos^{\Omega-2}\theta}{1\pm\cos^\Omega\theta}\right)\nonumber\,.
  \end{eqnarray}
In contrast with (\ref{rpq2}), this state leads to a small but {\it finite} fermionic concurrence 
\begin{equation}
    C_\pm(\theta)=\sin^2\theta\frac{\cos^{\Omega-2}\theta}{1\pm\cos^\Omega\theta}
    \label{Cexp}\,,
\end{equation}
which is  parallel (antiparallel) for positive (negative) $S_z$-parity. Nevertheless, it is  very small unless $\Omega$ and $\theta$ are sufficiently small. In a projected after variation treatment,  $\theta=0$ for $v_x<\varepsilon$ and (\ref{Cexp}) implies an appreciable concurrence just in a very narrow interval after the phase transition at $v_x=\varepsilon$, where $\theta$ is still small but non-zero. 

\begin{figure}
    \centering
    \includegraphics[scale=0.825]{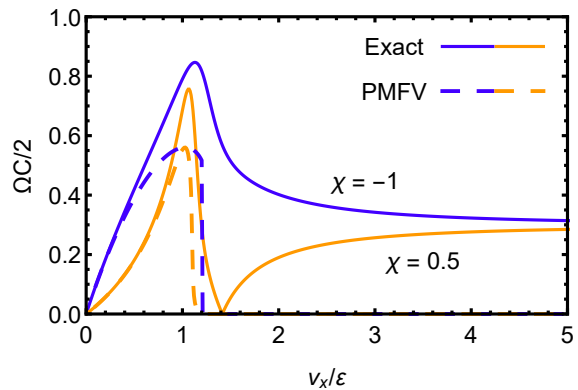}
    \caption{Comparison between exact results for the fermionic concurrence for $\chi=0$ and $\chi=0.5$ with $\Omega=50$, and those obtained with an $S_z$-parity projected (before variation) mean-field approach (PMFV).} 
    \label{fig5}
\end{figure}

For $v_x<\varepsilon$, it is possible to improve previous result by projecting {\it before} variation, i.e., by determining $\theta$ (and $P_z$) through the minimization of $\langle \Psi^\pm|H|\Psi^\pm\rangle$. In this case $\theta$ will be non-zero  $\forall$ $v_x>0$ and (\ref{Cexp}) will lead to a finite appreciable  concurrence in the normal  region ($0<v_x\alt\varepsilon$), where $\theta$ remains small. 
As seen in Fig.\ \ref{fig5}, the ensuing results are  accurate for small $v_x/\varepsilon\ll 1$, although  for  $v_x>\varepsilon$,  $\theta$  increases and the concurrence  (\ref{Cexp})  becomes again vanishingly small for appreciable $\Omega$, as $\rho_{pq}$ becomes essentially  a  convex mixture of Gaussian states. 

On the other hand, the up-down negativty  determined by the elements (\ref{elc}), 
\begin{equation}
    {\cal N}^{+-}_{\pm}=\frac{\sin^2\theta}{2(1\pm\cos^\Omega\theta)}\,,\label{neg5}
\end{equation}
is very close to the previous result (\ref{neg2}) derived from (\ref{rpq2})  for appreciable  $\Omega$ and $\theta$, thus providing a correct description for $\chi<1$. 

A final comment is that the projected states (\ref{factorpar0}) constitute the {\it exact} GS side-limits at the separability point (\ref{fac}) $\forall$ $\Omega$.  As we have seen, these states lead to a finite one-body entropy $E$ (and also finite  up-down entropy $E^{+-}$ and negativity ${\cal N}^{+-}$), 
which does not vanish for large $\Omega$, in contrast with the associated fermionic concurrence (\ref{Cexp}).  Thus, for not too small $\Omega$ this point will be clearly visible only in the concurrence, as verified in Fig.\ \ref{fig1} (see also Appendix). 

\subsubsection{Mean-field plus Random-phase approximation}
In order to obtain a reliable analytic description of the concurrence for large $\Omega$,  it is necessary to employ at least  a random-phase approximation (RPA) \cite{RS.80}, equivalent to a first-order Holstein-Primakoff bosonization of the Hamiltonian around the MF solution (see \cite{RS.80} and also  \cite{VPM.04,DV.05,BDV.06} and \cite{MRC0.08, MRC.10}). It can be implemented in a simple way  by mapping approximately the collective spin operators to boson operators. For  $0<v_x<\varepsilon$ (and $|\chi|<1$) this mapping takes the form 
\begin{eqnarray*}
    S_+\rightarrow \sqrt{\Omega}
    b^\dag,\quad S_-\rightarrow\sqrt{\Omega}
    b,\quad S_z\rightarrow b^\dagger b-\Omega/2\,,
\end{eqnarray*}
where the condition  $[b,b^\dag]=1$ is imposed. It is valid when  $\langle S_z\rangle$ is close to $-\Omega/2$, i.e. in the normal MF phase $v_x<\varepsilon$, where the vacuum $|0\rangle$ of $b$ corresponds to the unperturbed GS $|\Psi_0\rangle$.  This leads to an approximate quadratic boson Hamiltonian: 
\begin{eqnarray}
    H-E_0
\rightarrow H_b&=&(\varepsilon-w)b^\dag b-\frac{v}{2}({b^\dag}^2+b^2)\nonumber\\
&=&\lambda {b'}^\dag b'-\frac{\varepsilon-w-\lambda}{2}\,,\end{eqnarray}
where $E_0=\langle \Psi_0|H|\Psi_0\rangle$,   $w=W\Omega$, $v=V\Omega$, and $b'=\alpha b-\beta b^\dagger$ is the final normal boson operator,  with $(^\alpha_\beta)=\sqrt{\frac{\varepsilon-w\pm\lambda}{2\lambda}}$ 
($[b',{b'}^\dag]=\alpha^2-\beta^2=1$) and 
\begin{equation}\lambda=\sqrt{(\varepsilon-w)^2-v^2}=\sqrt{(\varepsilon-v_x)(\varepsilon-v_y)}\,,\end{equation}
the normal boson energy, where $v_y=\chi v_x$.  
The GS $|\Psi\rangle$ is then approximated by the vacuum $|0'\rangle=\alpha\,\exp[\frac{\beta}{2\alpha} {b^\dag}^2]|0\rangle$ of $b'$, 
which leads to $\langle b^2\rangle=\alpha\beta=\frac{v}{2\lambda}$, $\langle b^\dagger b\rangle=\beta^2$ and hence to 
$\langle S_+\rangle^2\approx \frac{\Omega v}{2\lambda}$, $\langle S_z\rangle\approx \frac{\varepsilon-w-\lambda}{2\lambda}-\Omega/2$. 
 With these values, 
we can estimate the concurrence through Eq.\ (\ref{CC}), 
which  for large $\Omega$ leads to the  asymptotic expression 
\begin{equation}C\approx \frac{1-\lambda/(\varepsilon-v_y)}{\Omega-1}\,,\;\;0<v_x<\varepsilon\,.\label{Crpa1}
\end{equation}
This value corresponds to a parallel-like concurrence (normal phase) \cite{DV.05,MRC0.08}.

\begin{figure}
    \centering
    \includegraphics[scale=0.725]{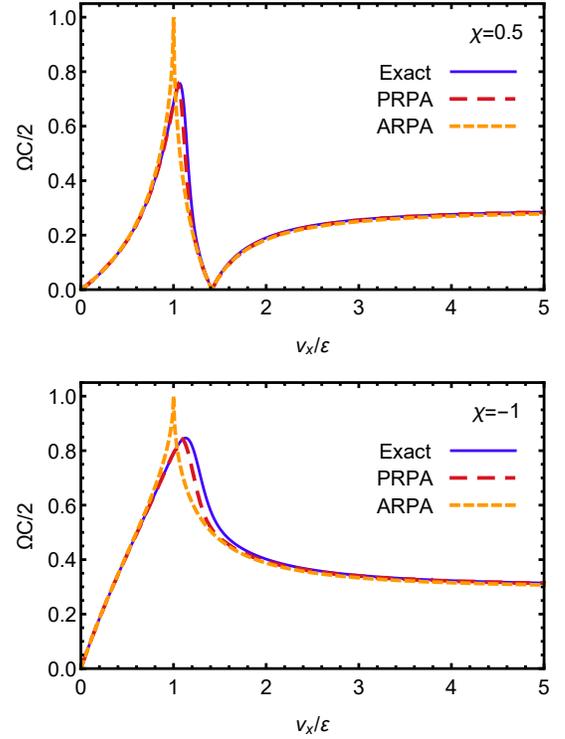}
    \caption{Comparison between exact and RPA results for the fermionic concurrence at two different anisotropies  and $\Omega=50$ fermions. Results from both the analytic expressions (\ref{Crpa1}), (\ref{Crpa2}) (ARPA), valid for large $\Omega$, and the finite $\Omega$ results derived from the states  (\ref{RPA1})--(\ref{RPA2}) (PRPA), are depicted.}
    \label{fig6}
\end{figure}

For $v_x>\varepsilon$, the bosonization should be done around  the parity breaking MF state  $|\Psi_{\rm mf}\rangle$ of Eq.\ (\ref{mfs}), which will correspond to the initial boson vacuum, and applied to the rotated operators $S'_{\pm}$, $S'_z$.  This leads to 
\begin{eqnarray}
    H-E'_0
\rightarrow H_b&=&(\varepsilon'-w')b^\dag b-\frac{v'}{2}({b^\dag}^2+b^2)\nonumber\\
&=&\lambda' {b'}^\dag b'-\frac{\varepsilon'-w'-\lambda'}{2}\,,\end{eqnarray}
where $E'_0=\langle\Psi_{\rm mf}|H|
\Psi_{\rm mf}\rangle$, $\cos\theta=\varepsilon/v_x$, 
$\varepsilon'=\varepsilon\cos\theta$, 
$w'=(v_x(3\cos^2\theta-2)+v_y)/2$, 
$v'=(v_x\cos^2\theta-v_y)/2$ and 
\begin{equation}\lambda'=
|\sin\theta|\sqrt{v_x(v_x-v_y)}\,.\end{equation}
The final expression for the asymptotic concurrence is 
\begin{eqnarray}C\approx \left\{\begin{array}{lcl}\frac{1-\lambda'/(v_x-v_y)}{\Omega-1}&,&\varepsilon<v_x<\varepsilon/\sqrt{\chi}\\\frac{1-(v_x-v_y)/\lambda'}{\Omega-1}&,&v_x>\varepsilon/\sqrt{\chi}
\end{array}\right.\label{Crpa2}\end{eqnarray}
where the upper (lower) formula corresponds to the parallel (antiparallel) concurrence, with the transition between both taking place at precisely the  separability point \eqref{fac} \cite{MRC0.08,RCM.08}. 
At this point $v'=0$ and  $\lambda'=v_x-v_y$, implying a zero RPA concurrence, in agreement with the exact result for large $\Omega$. Moreover, at this point $\varepsilon'-w'=\lambda'$ and $b'=b$, so that the bosonic vacuum remains unaltered, i.e., it is the MF GS. Thus, separability  is directly and exactly detected in  RPA. For $-v_x<v_y<0$ (i.e. $-1<\chi<0$) the concurrence is parallel and just the upper row in (\ref{Crpa2}) should be applied ($\forall$ $v_x>\varepsilon$). 
Expressions (\ref{Crpa1})--(\ref{Crpa2}) yield an excellent description of the fermionic concurrence for all $-1\leq \chi<1$ and  both $v_x<\varepsilon$ and $v_x>\varepsilon$ for appreciable $\Omega$,  as verified in Fig.\ \ref{fig6}. 

We finally mention that it is possible to further improve this approach for finite systems  by considering the fermionic state corresponding to the bosonic vacuum $|0'\rangle$. 
For $v_x<\varepsilon$, such state has the form  
\begin{equation}|\Psi_{\rm RPA}\rangle\propto\exp[\gamma S_+^2]|\Psi_0\rangle\label{RPA1}\,,\end{equation}
where $S_+=\sum_{p}c^\dagger_{p+}c_{p-}$. The parameter $\gamma$ can be taken from RPA ($\gamma=\frac{\beta}{2\Omega \alpha}$) or determined variationally. It is then seen explicitly that (\ref{RPA1}) has definite $S_z$-parity and  is not a SD,  containing genuine fermionic correlations. 

For $v_x>\varepsilon$  we may just replace $|\Psi_0\rangle\rightarrow|\Psi(\theta)\rangle$ and $S_+\rightarrow S'_+$ in (\ref{RPA1}), with $|\Psi(\theta\rangle)$ the MF state \eqref{mfs}, leading to 
 a state 
$|\Psi_{\rm RPA}(\theta)\rangle$. Moreover, in this case we may also consider $S_z$-parity restoration, which leads to 
 a projected RPA state 
\begin{equation}|\Psi_{\rm RPA}^\pm\rangle\propto |\Psi_{\rm RPA}(\theta)\rangle\pm|\Psi_{\rm RPA}(-\theta)\rangle\,.\label{RPA2}\end{equation} Results obtained in this way are very accurate for all $v_x$ and sizes, as verified in Fig.\ \ref{fig6}, but nevertheless approach the values (\ref{Crpa2}) for large $\Omega$. And for $\chi=1$, $S_z$ projection should be instead implemented, which in this model will directly lead to the exact GS's $|K\rangle$. 

\section{Conclusions}
\label{III}
We have analyzed the fermionic entanglement in the exact GS of the Lipkin model and its relation with bipartite entanglement. A global measure of fermionic entanglment such as the one body entropy, which measures the minimum relative entropy to a fermionic Gaussian state (thus vanishing iff  the GS is a Slater Determinant) and which can be accessed through one body measurements, correlates essentially with the mean-field order parameter associated with parity breaking,  increasing monotonously with the  coupling strength and saturating for strong couplings. It is also approximately proportional to  the total bipartite entanglement entropy between the  upper and lower modes, a quantity which has no physical analogue in a pure spin realization of the model. Both entropies scale with the size $\Omega$ of the system, are weakly dependent on the anisotropy and can be correctly (and analytically) described  by an $S_z$-parity breaking mean-field approach with basic symmetry restoration. This symmetry plays an important role in the previous proportionality and ensures a fixed number parity in the entangled up or down modes, enabling to consider the total up-down entanglement as a standard quantum resource.  

We have also analyzed the fermionic entanglement of the reduced state $\rho_{pq}$ of two up-down pairs (first non-trivial case) through the fermionic concurrence. As opposed to previous global measures, 
this concurrence, which measures the  deviation of $\rho_{pq}$ from a convex mixture of Gaussian states (i.e., from a statistical mixture of Slater Determinants)  presents a dominant peak at the phase transition,  whose sharpness 
increases for $\chi\rightarrow 1$ (where the maximum  corresponds to the narrow interval where the $|K=1\rangle$ eigenstate is the GS). While this concurrence can be directly associated with a bipartite mode entanglement (i.e.\ to an effective $p-q$ spin $1/2$ pair), it differs from the up-down mode entanglement of $\rho_{pq}$,   which can be measured through the pertinent negativity and behaves similarly to previous global measures. 
The mean-field approach can  correctly predict this  negativity through basic symmetry restoration, but it cannot  yield a non-vanishing concurrence for strong couplings even after full $S_z$-parity projection. Only by  considering RPA (collective-boson type) correlations on top of the mean-field it is possible to capture its proper behavior. The present study then provides a detailed analysis of the fermionic entanglement in a  strongly interacting system and its potential for quantum information through its relation with relevant bipartite entanglement measures.  

\acknowledgments
The authors acknowledge support from CONICET (MDT,NG) and  CIC (RR) of Argentina. MC was supported by the Center for Nonlinear Studies and the Laboratory Directed Research and Development (LDRD) program at Los Alamos National Laboratory. Work supported in part by CONICET PIP  112201501-00732.

\appendix
\section{Fermionic separability \label{ferfact}}
The exact definite $S_z$-parity GS (\ref{GSX}) of the Lipkin model is entangled,  i.e.\ it is not a SD,  
if $|S_z|<\Omega/2$. However, at the parity transitions arising for increasing $v_x/\varepsilon$ at fixed $\chi\in(0,1)$, the GS becomes two-fold degenerate and exact GS's with no definite $P_z$ become feasible \cite{RCM.08}. In particular, at the {\it first} transition (Eq.\ (\ref{fac})) an $S_z$-parity breaking SD of the form (\ref{mfs}) (the MF GS)  becomes an {\it exact} GS for any $\Omega$ (fermionic separability). 

This result can be easily seen by writing the Hamiltonian  of Eq.\ (\ref{H2})  in terms of the rotated quasispin operators $S'_{\pm}=\sum_p {c'}^\dag_{p\pm}c'_{p\mp}$, 
$S'_z=\frac{1}{2}\sum_p ({c'}^\dag_{p+}c'_{p+}-{c'}^\dag_{p-}c'_{p-})$, determined by the primed fermion operators (\ref{cp}). We obtain, dropping the constant term $V_x(1+\chi)N/4$, 
\begin{equation}
   \begin{array}{rcl} H&=&\varepsilon\cos\theta S'_z-V_x({S'_x}^2\cos^2\theta+{S'_z}^2\sin^2\theta+\chi {S'_y}^2)\\
    &&+\sin\theta[\varepsilon S'_x+V_x\cos\theta (S'_x S'_z+S'_zS'_x)]\,.
    \end{array}
    \label{r12}
\end{equation}
The MF stationary condition (\ref{ct}) cancels the single fermion excitation terms $\propto S_+'$, present in the second line of (\ref{r12}), when applied to the MF state (\ref{mfs}) (for $\bm{n}\cdot\bm{S}=S_y$). And the two-fermion excitation terms $\propto {S_+'}^2$  in the first line of (\ref{r12}) also vanish iff the condition 
 \begin{equation}\cos\theta=\sqrt{\chi}\label{ctf}\end{equation}
 is satisfied, in which case the MF state (\ref{mfs})--(\ref{cp}) becomes an exact eigenstate, obviously for both signs of $\theta$.   Since for $\theta>0$ its coefficients $C_K(\theta)$, Eq.\ (\ref{CKt}), are all positive, such state cannot be orthogonal to the GS (\ref{GSX}) (where $C_K\geq 0$ $\forall$ $k$) and must then be a GS. Eq.\ (\ref{ctf}) then leads, together with \eqref{ct}, to the factorizing point  (\ref{fac}). This separability is  equivalent to that occurring in the spin representation \cite{RVF.04,RCM.08,AA.06,GAI.08}, where it corresponds to a factorized (i.e., product) GS. 

The {\it side limits} of the exact GS at this transition do have, however, a definite parity  and are then given by the $S_z$-parity projected MF states (\ref{factorpar0}) \cite{RCM.08}, with $\theta$ determined by (\ref{ctf}). These states are both entangled for $\theta\in(0,\pi)$, i.e., $\chi<1$ (for $\chi\rightarrow 1$, $\theta\rightarrow 0$ and $|\Psi^{+}\rangle\rightarrow |\Psi_0\rangle=|K=0\rangle$, whereas   $|\Psi^-\rangle\rightarrow|K=1\rangle$). 
Therefore, at the  side-limits  of the separability  point,  the exact one body entropy approaches the values  given  by Eq.\ (\ref{SspP}), i.e., essentially the diagonal MF result  (\ref{Ssptheta}) for not too small $\theta$ and $\Omega$, which are {\it non-zero} and the same for  both parities (the same occurs with the side-limits of $E^{+-}$ and ${\cal N}^{+-}$, 
 Eqs.\ (\ref{pmfsup}) and (\ref{neg5})). In contrast,  the exact side-limits  (\ref{Cexp}) of the four-level fermionic concurrence become very small for not too small $\theta$ and $\Omega$, as $\rho_{pq}$ is essentially a convex mixture of Gaussian states.
Consequently, the existence of separability   at an $S_z$-parity transition is clearly reflected in the vanishing value of the fermionic concurrence, but not in the behavior of $E$ (nor $E^{+-}$ or ${\cal N}^{+-}$),  as seen in Fig.\ \ref{fig1}.

%

\end{document}